# Measuring the speed of light and the moon distance with an occultation of Mars by the Moon: a Citizen Astronomy Campaign

Jorge I. Zuluaga[1,2,3,a], Juan C. Figueroa[2,3], Jonathan Moncada[4],
Alberto Quijano-Vodniza[5], Mario Rojas[5], Leonardo D. Ariza[6],
Santiago Vanegas[7], Lorena Aristizábal[2], Jorge L. Salas[8],
Luis F. Ocampo[3,9], Jonathan Ospina[10], Juliana Gómez[3], Helena Cortés[3,11]

[2] *FACom - Instituto de Física - FCEN, Universidad de Antioquia, Medellín-Antioquia, Colombia*
[3] *Sociedad Antioqueña de Astronomía, Medellín-Antioquia, Colombia*
[4] *Agrupación Castor y Pollux, Arica, Chile*
[5] *Observatorio Astronómico Universidad de Nariño, San Juan de Pasto-Nariño, Colombia*
[6] *Asociación de Niños Indagadores del Cosmos, ANIC, Bogotá, Colombia*
[7] *Organización www.alfazoom.info, Bogotá, Colombia*
[8] *Asociación Carabobeña de Astronomía, San Diego-Carabobo, Venezuela*
[9] *Observatorio Astronómico, Instituto Tecnológico Metropolitano, Medellín, Colombia*
[10] *Sociedad Julio Garavito Armero para el Estudio de la Astronomía, Medellín, Colombia*
[11] *Planetario de Medellín, Parque Explora, Medellín, Colombia*

## ABSTRACT

In July 5th 2014 an occultation of Mars by the Moon was visible in South America. Citizen scientists and professional astronomers in Colombia, Venezuela and Chile performed a set of simple observations of the phenomenon aimed to measure the speed of light and lunar distance. This initiative is part of the so called "Aristarchus Campaign", a citizen astronomy project aimed to reproduce observations and measurements made by astronomers of the past. Participants in the campaign used simple astronomical instruments (binoculars or small telescopes) and other electronic gadgets (cell-phones and digital cameras) to measure occultation times and to take high resolution videos and pictures. In this paper we describe the results of the Aristarchus Campaign. We compiled 9 sets of observations from sites separated by distances as large as 2,500 km. We achieve at measuring the speed of light in vacuum and lunar distance with uncertainties of few percent. The goal of the Aristarchus Campaigns is not to provide improved values of well-known astronomical and physical quantities, but to demonstrate how the public could be engaged in scientific endeavors using simple instrumentation and readily available technological devices. These initiatives could benefit amateur communities in developing countries increasing their awareness towards their actual capabilities for collaboratively obtaining useful astronomical data. This kind of exercises would prepare them for facing future and more advanced observational campaigns where their role could be crucial.

## 1. Introduction

Eclipses and occultations are common astronomical phenomena. They have inspired astronomers in

a   Corresponding author: jorge.zuluaga@udea.edu.co

*Last update: 01/06/15*

the past to use them as ways to measure the size of the local Universe and several physical constants. Well known are the methods devised originally by Aristarchus of Samos (c. 310 – c. 320 BCE) and Eratosthenes of Cyrene (c. 276 B.C. - c. 195/194 B.C.) who provided the first precise estimations of the size of our planet and the distance and size of the Moon and the Sun, using for that purposes the phases of the moon, solar and lunar eclipses and the shadow cast by the columns of a temple[1,2,3].

More recently physical constants such as the gravitational constant and the speed of light have attracted the attention of astronomers who have devised ingenious methods to measure these contants using the huge temporal and spatial scales available in Astronomy. Such is the case of the Ole Roemer (1644-1670) who devised a method to measured the speed of light in vacuum observing the periodic occultations of the moon Io by its host planet, Jupiter.[4].

Many centuries afterwards, astronomers have measured such quantities to exquisite precision. The radius of our planet for instance has been measured to a precision of millimeters[5]. The distance to the Moon has been measured to a precision of centimeters using lasers and arrays of mirrors placed on the Lunar surface by Apollo Astronauts[5,6]. The size and distance of the Sun have been determined to precisions between $10^{-9}$ and $10^{-11}$ by different astronomical methods including Venus and Mercury transits and more recently using advanced techniques of radar ranging and celestial mechanics[8,9,10]. On the other hand, the speed of light, that until few decades ago was too high for laboratory precise measurements and hence required astronomical methods, is now regularly measured with precisions of the order of $10^{-9}$ via the determination of wavelengths and frequencies of electromagnetic waves in Earth laboratories[11].

As opposed to ancient or historical methods, most of those modern techniques are far from the realm and technological capacities of amateur astronomers and non-scientists. The measurement of the local Universe and other physical constants seems to be a matter of sophisticated instrumentation and advanced scientific capabilities. This is contrary to the spirit of the relatively simple measurements proposed originally by pioneers such as Aristarchus, Eratosthenes or Roemer.

Historical methods used to measure the local Universe are common place in Astronomy and Physics textbooks[12]. Moreover, many efforts to use them as educational tools in the classroom have also been devised and published[13-18]. Today a plethora of advanced and accessible technological devices such as smartphones, tablets, digital cameras and precise clocks, are opening a new door to the realm of "do-it-yourself-science". Almost all of these personal electronic devices come today with a GPS receiver, fast internet connections and precise clocks synchronized to international time systems. This technological state-of-the-art is giving us incredible opportunities to involve science enthusiasts and in general in global scientific projects, where their role is to provide in situ measurements that will be otherwise prohibitively expensive for a professional scientific team. This scientific and sociological phenomenon is called "Citizen Science"[19]. A recent example of the power of citizen science arose in 2013 when hundreds of pictures and videos of the Chelyabinsk Impact Event taken by casual observers of the phenomenon and available in YouTube and other internet sites allowed several scientific groups around the world to study the phenomenon with unprecedent detail[20-23]. Other well-known project is the "Globe-at-Night" project[24], a successful initiative intended to measure light pollution all around the world.

Observational campaigns, where many amateur and professional astronomers and sometimes casual observers, register an astronomical phenomenon from many different places and times on Earth, are not new[25]. Particularly interesting have been for example the observational campaigns to follow the evolution of comets, measuring with precision the position of asteroids or observing the occultation of stars by solar system objects[26]. In most cases, however, being involved in these campaigns require a lot of experience and advanced observational or photographical skills. Most simple observational campaigns, aimed to measure or observe well studied phenomena and involving simpler measurements and more readily available instrumentation, may work as an entry point to enthusiasts ager to participate on more advanced projects. Moreover, a massive participation on these simple campaigns may enhance the preparation of a wide area community to face more interesting and challenging observational collaborative observations.

With all these in mind we launched in Colombia in 2014, a citizen science campaign we dubbed the "Aristarchus Campaign"[27]. The Aristarchus Campaign, that completes today 3 different versions starting with a first one on 15 April 2014 and that was intended to observe and measure the Lunar Eclipse on the same date[28], was aimed this time to measure the Occultation of Mars by the Moon on 5 July 2014. The occultation was visible with very favorable conditions across the North part of South America (see Figure 1).

In this paper we present a description of the campaign (Section 2) including the measurements that was performed and the information obtained from them (Section 3 and 4). We also present and summarize the best observations reported by a community of enthusiast, amateur and professional astronomers who participated in the campaign (Section 5). The scientific results derived from the analysis of those observations are also described in detail (Section 6). We have achieved at measuring the speed of light with an uncertainty of ~7% and the lunar distance with a maximum uncertainty of 2%. The limitations, suggestions, conclusions and future prospects are also presented (Section 7). The experience described here and the derived results could be useful to organize and execute future similar initiatives worldwide.

## 2. The Aristarchus Campaign

Originally motivated by the Lunar Eclipse of 15 April 2014, the Aristarchus Campaign is a collaborative initiative aimed at making the experience of observing frequent astronomical phenomena (eclipses, occultations, meteor showers, etc.), not only enjoyable and recreative activities, but also great opportunities to perform simple measurements with readily available astronomical equipment (binoculars, small telescopes, cameras) and personal technological devices (cell-phones, electronic tablets). In most cases those measurements can be used to reproduce historical astronomical observations or to calculate astronomical and physical constants for educational purposes.

At the date of writing there have been 3 versions of the Aristarchus Campaign. The first version of the Campaign was related to the Lunar Eclipse of 15 April 2014 and succeed at motivating the community to measure several properties of the Moon and the Lunar Eclipse during that night. Regretfully, the weather in Colombia (where the Campaign was initially launched) made impossible to perform most of the interesting measurements during the Eclipse itself. However a photographic record of the lunar ascend was performed by one citizen scientist and allows us to reliably measure the variation in the lunar apparent diameter. From these measurements we were able to determine the lunar distance within an uncertainty of less than 3%[28].

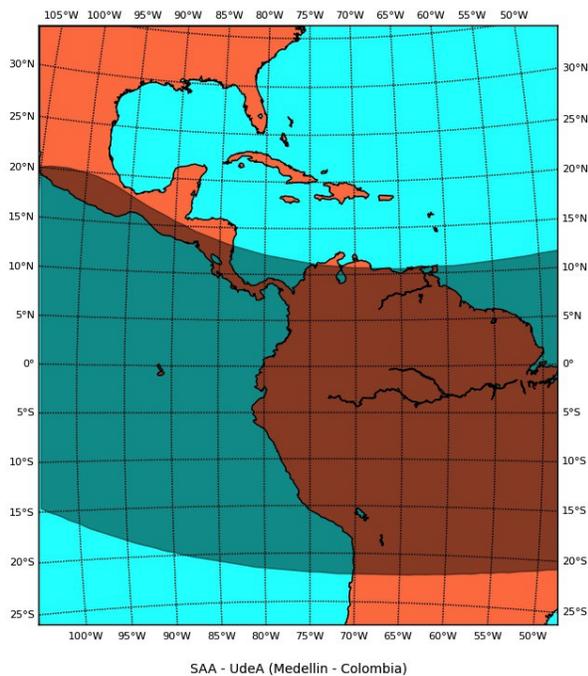 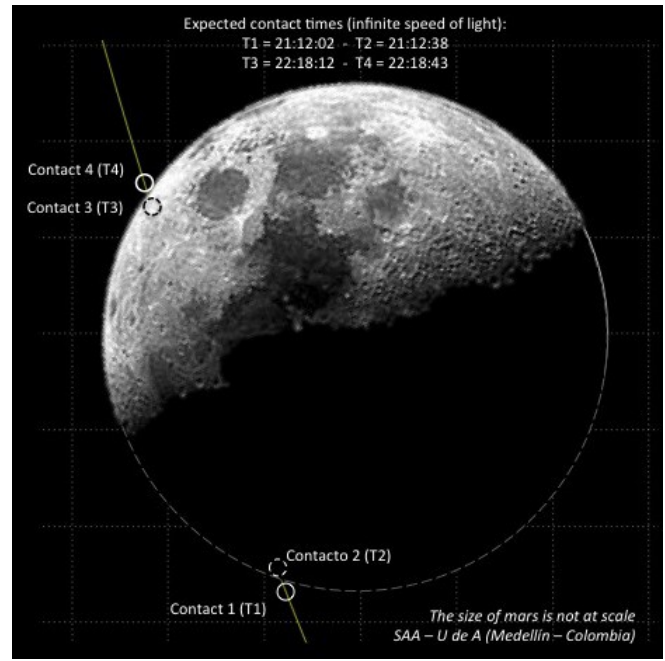

*Figure 1. Left panel: zone of Totality of the Occultation of Mars by the Moon in 5 July 2014. The shaded area encloses the region in South America where observers could contribute with observations and pictures for the Aristarchus Campaign. Right panel: artistic diagram showing the phases of the occultation as observed from Bogotá, Colombia. The occultation started (contacts 1 and 2) when Mars disappeared behind the dark side of the lunar disk.*

The occultation of a bright planet by the Moon is a perfect opportunity for motivating the community of amateur astronomers and science enthusiasts in a wide geographic area, to perform simple measurements of the phenomenon. In this case we use social networks and e-mail lists to invite people located in the North part of South America to measure visually or photographically the contact times of the occultation of Mars by the Moon and/or to take time-lapses and videos of the ingress and egress of Mars.

A particularly advantageous feature of this occultation was that the ingress of Mars behind the Moon occurred in the dark side of the Lunar disk (see right panel in Figure 1). This feature allowed precise visual determinations of the time of Contact 2 (complete disappearance of the Martian disk) without requiring sophisticated astronomical equipment or pictures of the phenomenon. Some of the observations that was reported by the enthusiasts in this campaign were made using binoculars and a precisely synchronized clock and still they were good enough for the purpose of this work.

Few weeks before the event we published a simple guide for the observation of the event[27]. Then few days before we published a detailed guide intended for more advanced amateur astronomers and for professional. In those guides we provide details about which specific measurements should be performed and how to achieve the required precision for the purposes of the campaign. We also developed and published a simple on-line application able to provide the expected contact times for any location[29]. This application was particularly useful for the campaign since other sites and free software intended for similar purposes (e.g. Occult[30]) are very specialized and tend to confound the casual enthusiasts. Moreover most of those alternative sites provided the expected contact times with the light-time correction, while, for our purposes, we needed to calculate times without this correction. This feature was particularly important to highlight the delay between the expected and the observed time and hence the effect that a finite speed of light has in astronomical observations (see Section 4.1).

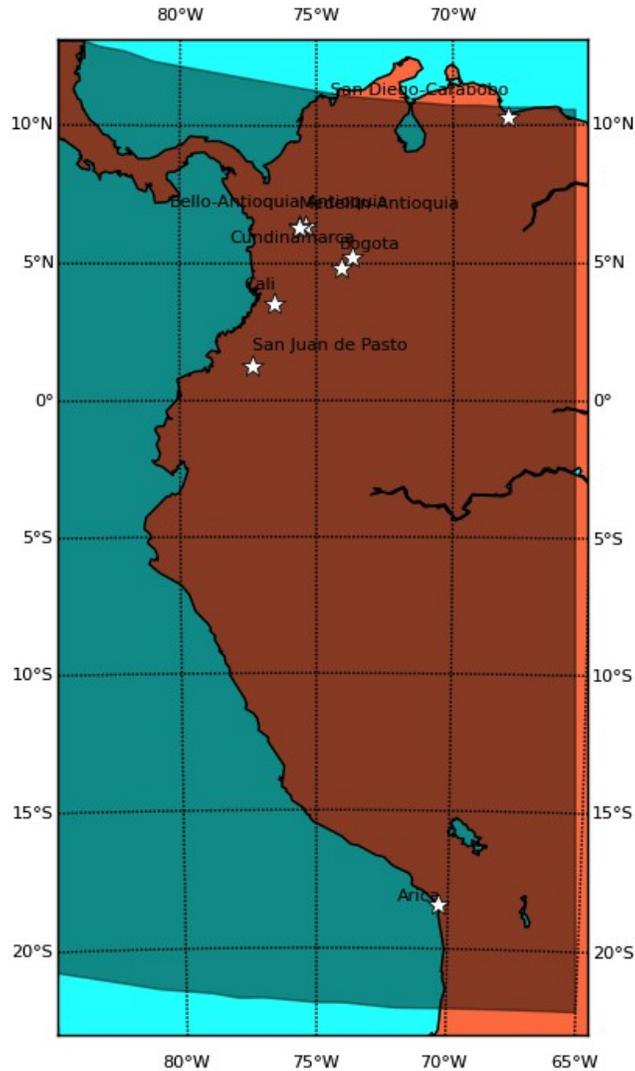

*Figure 2. Locations of the observers contributing to the Aristarchus Campaign of the Occultation of Mars by the Moon.*

9 groups and individual observers reported results from location across a wide geographic area covering Colombia, Venezuela and the north of Chile (see Figure 2). Most of the reports arrived or were discovered in social networks (Twitter and Facebook). Others were reported via e-mail. Full resolution images and videos in high quality formats such as fits, tiff and raw, were submitted using an internet public repository.

**3. Occultation geometry and calculation of the contact times**

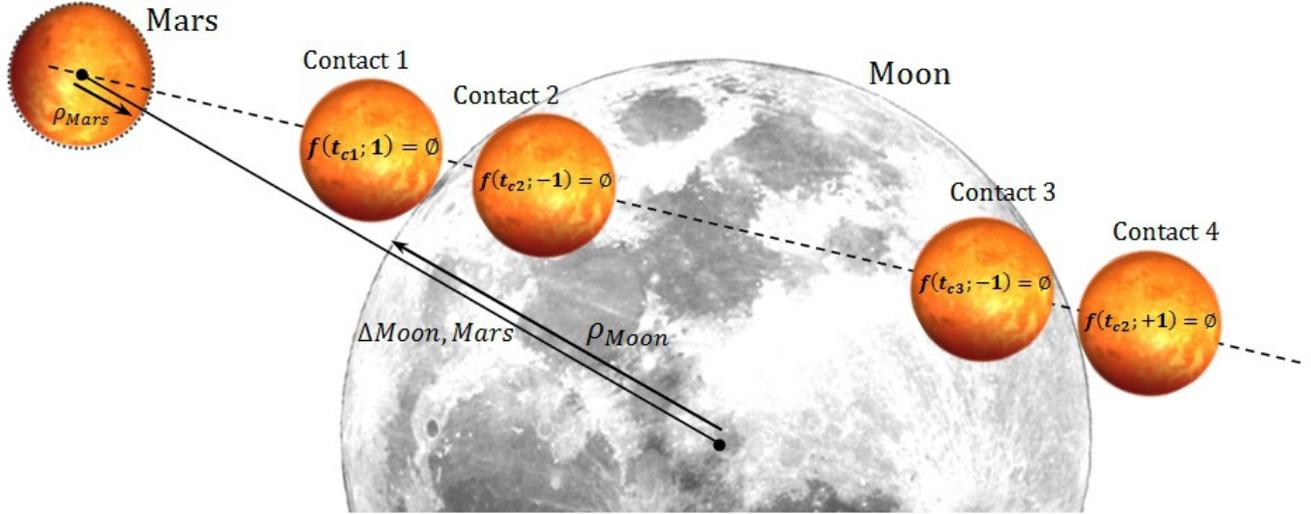

*Figure 3. Geometry of the occultation. The size of the Martian disk is exaggerated. The apparent trajectory followed by the disk of Mars between the first and last contact is called the occultation cord. The occultation function f(t,k) is defined by Equation (1).*

The occultation of a planet by the Moon happens when the angular distance between the centers of the Moon and the planet is smaller than the sum of their apparent diameters. More generally, we define an occultation as the period of time when the following function get negative values:

$$f(t;k) = \Delta_{Moon,Mars}(t) - \rho_{Moon}(t) - k\rho_{Mars}(t) \qquad (1)$$

Here $\Delta_{Moon,Mars}$ is the angular distance between the centers of the Moon and the planet (Mars in this case), $\rho_{Moon}$ and $\rho_{Mars}$ are their apparent radii and *k* is a constant that define the initial and final phases of the occultation (see Figure 3):

- Contact 1: The disk of the planet starts to be occulted ($k = +1$).
- Contact 2: The whole disk of the planet is occulted ($k = -1$).
- Contact 3: The disk of the planet starts to appear behind the lunar limb ($k = -1$).
- Contact 4: The occultation ends ($k = +1$).

Additionally we recognize two intermediate contacts: Contact 1.5 and contact 3.5 corresponding to the times when the center of the Martian disk is exactly over the limb of the Moon, i.e. approximately half of the Martian disk is occulted ($k=0$).

In order to compute the times of each contact we need to calculate the precise position in space (state vector) of Mars and the Moon at a given time with respect to an observer on a given location on the surface of the Earth. Using the state vector of the planet (as referred to the J2000.0 equinox) we calculate its celestial coordinates (right ascension and declination) and distance. The apparent radius of the Moon and the planet is calculated using their mean physical radius. With the celestial coordinates and apparent radius of the bodies we proceed at calculating the value of the function *f(t;k)* at any time t. We used a bisection algorithm to find the exact times when this function is zero (contact times).

For calculating the position in space of the Earth, Mars and the Moon we use the data and tools of the Navigation and Ancillary Information Facility (NAIF) of NASA, in particular the ANSI C API of the SPICE system[31]. In all cases we calculated planetary positions without any light-time correction. This procedure provide us with the contact times as expected if the speed of light was infinite. The positions were calculated with respect to the J2000.0 equinox. Precession and nutation only modifies the absolute coordinates of the objects, but not their relative positions that finally determines the contact times. The location of the observers in space and its transformation to the J2000.0 equinox was computed using the International Terrestrial Reference Frame 1993 (ITRF93).

## 4. Methods

### 4.1. Measuring the speed of light

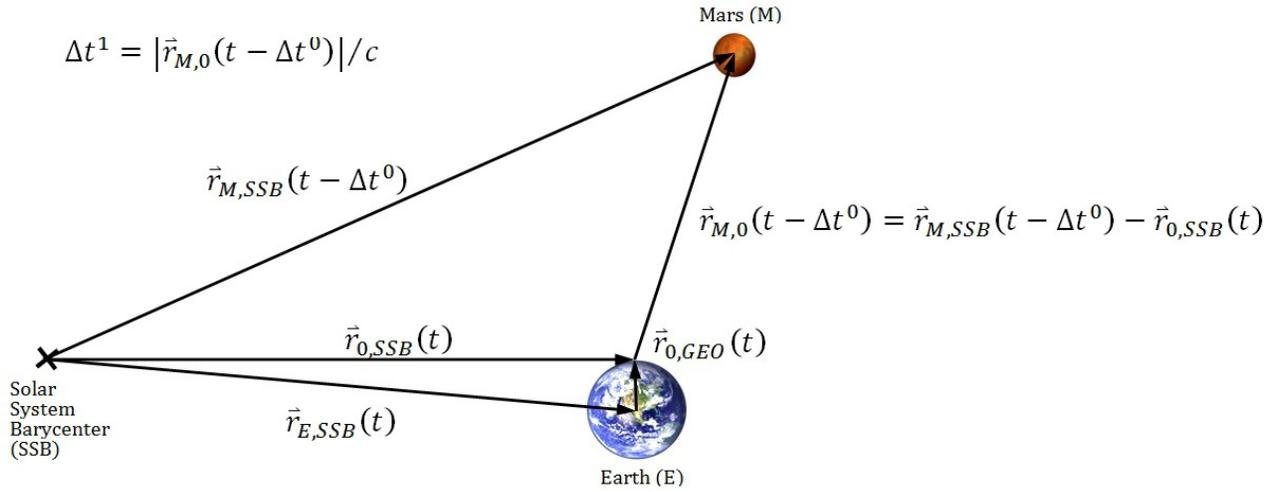

*Figure 4. Geometrical configuration of Mars and the Earth with respect to the Solar System Barycenter. Due to the finite speed of light, the observed position of Mars must be calculated obtained from its position in space a time Δt before the observation date t. The light-time Δt is computed using a numerical procedure further explained in the text.*

It is a matter of fact that when we observe an object in the sky we see the object as it was in the past. This effect is a by-product of the finite speed of light. For bodies with a substantial apparent motion (e.g. solar system objects) this light-time effect produces a detectable difference between its predicted and actual position. Neglecting relativistic effects, the light-time *Δt* of a given object (e.g. the Moon or Mars) can be calculated using the procedure illustrated in Figure 4.

Starting with a guess value of the light-time *Δt⁰* we calculate the position of the target body at a time *t* – *Δt⁰* with respect to an inertial frame of reference (Solar System Barycenter, SSB). The position of the Earth with respect to the SSB and the observer with respect to the center of the Earth is calculated at the observing time *t*. This procedure provides us an updated value of the distance between the body and the observer $|\vec{r}_{M,O}(t-\Delta t^0)|$. Using this result, an improved value of the light-time *Δt¹* can be obtained (see formula in Figure 4). The procedure is repeated until it fulfills a given convergence criterium. The final outcome of this numerical procedure is the precise light-time and the observed position of the target body in space.

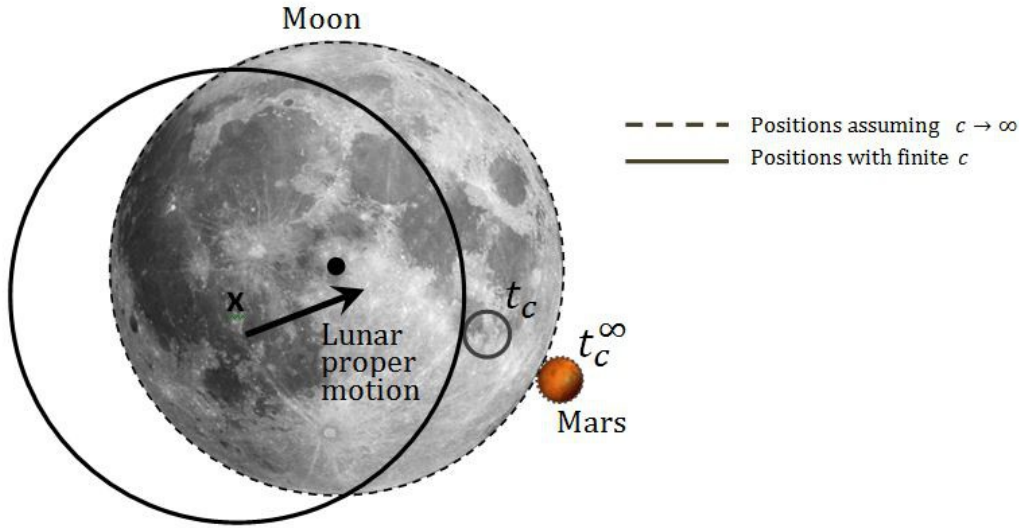

*Figure 5. Illustration showing the position of Mars and the Moon in the sky assuming an infinite speed of light, c (dashed lines and photographic representations) and with a finite value of c, i.e. their actual positions (solid lines). Due to the proper motion of the Moon and Mars the contact times in both cases, $t_C, t_C^\infty$ are different. Sizes and apparent motion were exaggerated for illustration purposes.*

For the date and time of the occultation, the light-time of the Moon and Mars were approximately 1.29 seconds and 510.0 seconds (8.50 minutes) respectively. At the same time their apparent motions in the sky were approximately ~21"/minute and ~1"/minute respectively (1" = 1 arc second). It implies that while the light of mars arrived to the Earth, its position in the sky was almost ~8.5" off its expected position (see Figure 5). In the case of the moon the light-time is responsible for a negligible difference of just ~0.3" between its actual and its expected positions. The resulting effect of the displacement in the apparent image of Mars due to the finite speed of light, is that the occultation starts ~8.5"/ (21"/minute) ~24 seconds earlier than expected if its image had arrived instantaneously (infinite speed of light). This is precisely the effect we expected to measure using the occultation.

**4.2. Measuring the lunar distance**

Observing an occultation offers us a unique opportunity to measure the distance to the Moon through the well known effect of parallax. Using Mars as a reference point, we can measure the angular displacement of the Moon when observed from different locations in the Earth

In the right panel of Figure 6 we show the occultation and the change in apparent position of the Moon with respect to Mars (Martian Reference Frame). The left panel show the occultation in the more convenient reference frame defined by the moon. The time of the initial and final occultation contacts (Contacts 1 and 4) is also shown and the position of the lunar center (left panel) and martian disk (right panel) are indicated over their apparent trajectories.

The parallax of the Moon is defined as the angular distance π between the centers of the lunar disk when observed at the same time by two observers. For convenience we have chosen the time of the last contact (Contact 4) as measured by the Observer 1, $t_{C4,1}$, as the reference time. The position of the lunar disk center as observed in the second location is marked with an empty circle over its corresponding trajectory at that particular location. It should be noticed that the parallax is not simply the distance between the trajectories of the lunar disk centers. The orientation of the line over which the parallax is measured (dashed line connecting contact 4 times in Figure 6) will depend on complex factors including the direction of the apparent motion of the moon and the difference between the geographic longitude of both observers.

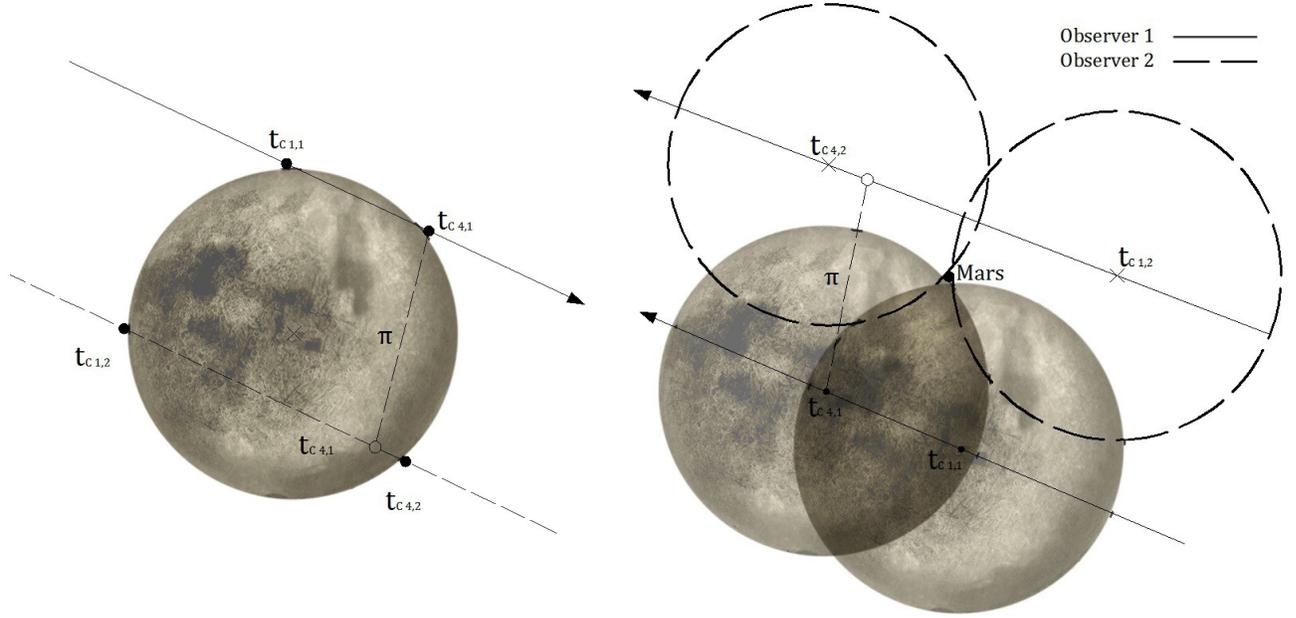

*Figure 6. Left panel: the occultation observed from two different locations as observed with respect to the lunar disk (conventional representation). Right panel: the same events as observed in a reference frame where Mars is at rest in the sky. The lunar pictures and the continuous lines indicate the disk and trajectory of the lunar center as observed from the location of Observer 1. Dashed lines correspond to observations performed at a second location. It can be noticed that the angle π measured directly over the pictures (lunar reference frame) correspond approximately to the parallax of the Moon which can only be strictly defined in a fixed reference frame (right panel).*

As schematically demonstrated by Figure 6 the parallax of the Moon is approximately given by the angular distance of the martian disk as measured at the same time by two observers.

In Figure 7 we show the same observations in 3-D as made from two different locations on the Earth surface. The distance B between the observers can be estimated from their precise geographical locations (longitude, latitude and altitude with respect to the see level). The cosine law relates $B$ and $\pi$ with the Moon distance to the Observers, $D_1$ and $D_2$:

$$B^2 = D_1^2 + D_2^2 - 2 D_1 D_2 \cos \pi \qquad (1)$$

Since $B \ll D_1, D_2$ and $D_1 \simeq D_2$, we can write $D_2 = D_1 + d$ where $d \ll D_1$. In terms of $D_1$ and $d$, Equation (1) can be written as:

$$B^2 = D_1^2 \left[ \frac{d^2}{D_1^2} + 4 \left(1 + \frac{d}{D_1}\right) \sin^2\left(\frac{\pi}{2}\right) \right] \qquad (2)$$

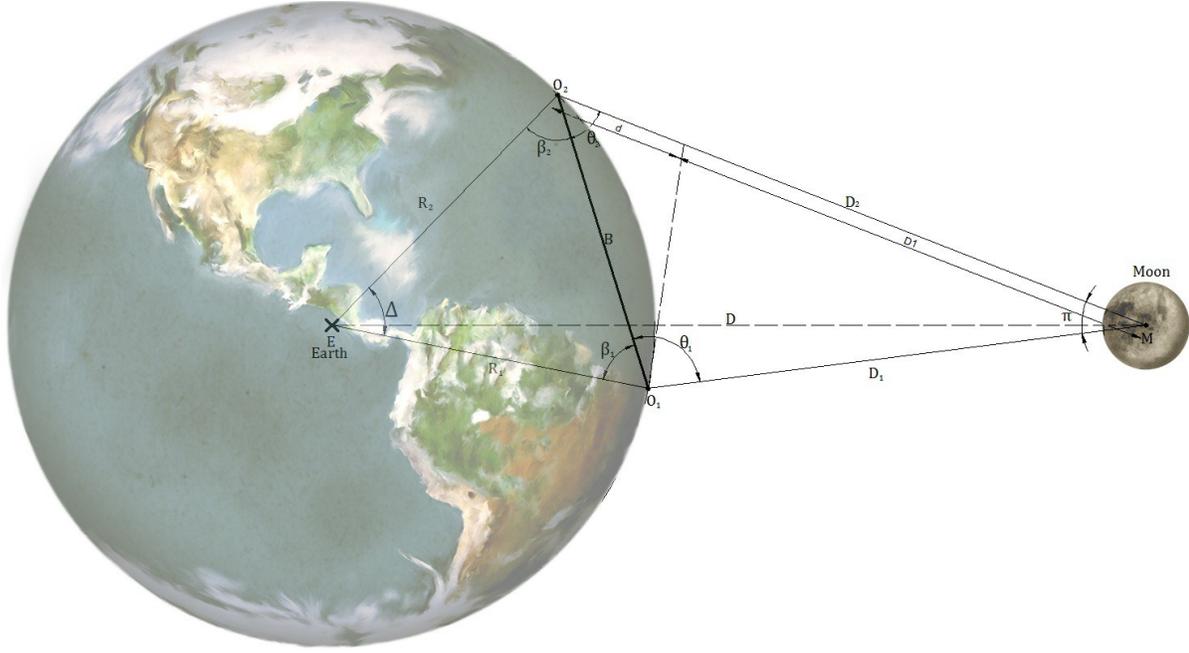

*Figure 7. Geometrical construction used to estimate the lunar distance D from the parallax π measured after observing the occultation from two locations $O_1$ and $O_2$. See Appendix 1 for detailed formulae.*

It should be stressed that the parallax angle $\pi$ must not be confounded with *pi* the mathematical constant. Neglecting in (2) second-order terms in $d/D_1$, we can write an approximate expression for $D_1$:

$$D_1 \simeq \frac{B}{2\sin(\pi/2)}\left(1 - \frac{d}{2D_1}\right) \quad (3)$$

Since the maximum value of $d$ is the Earth's radius $R_E$ and it can adopt positive or negative values depending on the relative positions of the observers, we can at first order constrain the value of $D_1$ in the range:

$$D_1 \in \frac{B}{2\sin(\pi/2)}\left(1 \pm \frac{R_E}{2D_1}\right) \quad (4)$$

With $D_1$~400,000 km and $R_E$~6,400 km, this expression provides the lunar distance with a minimum theoretical uncertainty of 1.6%, provided B and the parallax are exact. This implies that Equation (4) is only reliable if the experimental uncertainty in $D_1$, arising mainly from uncertainties in B and $\pi$, is not larger than ~2%.

On the other hand the geocentric lunar distance $D$ can be constrained using the triangle $MO_1O_2$. At first order in $R_E/D_1$ the geocentric distance will be in the interval:

$$D \in (D_1 + R_E)\left(1 \pm \frac{R_E}{2D_1}\right) \quad (5)$$

If we could determine independently the exact direction of the Moon center (as measured for example by its right ascension and declination), more precise values for $D_1$, $D_2$ and $D$ can be obtained. Appendix 1 present detailed formulas in this case.

**4.3. Using multiple sites to improve the lunar distance**

If instead of having just two locations we compile information from multiple vantage points, we can attempt at determining the lunar distance with an improved precision. For that purpose we can increase the number of free variables in our theoretical model, from one (the lunar distance) to three (the lunar distance and its celestial coordinates, right ascension, $\alpha$, and declination, $\delta$). The formula to calculate precisely the lunar distance from angles $\pi$, $\alpha$ and $\delta$ are presented in Appendix A.

But measuring $\alpha$ and $\delta$ in the context of a citizen campaign is an unrealistic goal. If instead of measuring these quantities we fit them as unknowns against at least 3 different values of the parallax, we can attempt at estimating the lunar distance with an improved precision. In Section 6 we show a comparison of the result of applying this multi-site fitting procedure with the first order approximation obtained with Equations (4) and (5).

**5. Observations**

We have compiled results from 9 different places across Colombia, Venezuela and Chile. In Table 1 we present a summary of the observations performed on each location along with detailed information of the places, observers and the methods used to measure geographic location and contact times.

*Table 1. Summary of observations provided by collaborators of the Aristarchus Campaign. The method to measure or estimate contact times are abbreviated as "Vi" (Visual), "Ex" (Extrapolated from a time-lapse before or after the occultation) and "Ph" (Photographic or photometric using pictures taken during the ingress or egress). Locations are sorted from south to north, according to geographic latitude. The method used to determine the geographic position is also indicated: "GPS" (GPS equipment or cell-phone, typical uncertainties of 10 meters) and "GIS" (Geographical Information Software, Google Earth[32], typical uncertainties of 100 meters)*

| # | Location<br>City<br>Country | Observer(s) /<br>Equipment | Lat.<br>Lon.<br>Alt. (m)<br>Method | Observations<br>Ingress (C1,C2), Egress (C3,C4)<br>Times in UTC |
|---|---|---|---|---|
| 1 | –<br>Arica<br>Chile | J. Moncada /<br>Telescopio Celestron<br>NexStar 130 SLT.<br>Cámara DSLR<br>Canon XS | -18.435000<br>-70.293333<br>24<br>GPS | **Ingress**:<br>*Time lapse* / All Times (Ph.)<br>**Egress**:<br>*Time lapse* / All Times (Ph.) |
| 2 | Observatorio de Nariño<br>San Juan de Pasto<br>Colombia | A. Quijano-Vodniza<br>M. Rojas /<br>CGE Pro 1400<br>CELESTRON 14" /<br>STL -1001E SBIG | +1.212222<br>-77.290833<br>2500<br>GPS | **Ingress**:<br>(No images) / (No times)<br>**Egress**:<br>*Time lapse* / All Times (Ex.) |
| 3 | –<br>Cali<br>Colombia | L. Aristizabal /<br>Visual Observations<br>7x50 Binoculars | +3.486745<br>- 76.489949<br>1000<br>GIS | **Ingress**:<br>(No images) / Time C1.5 (Vi.)<br>**Egress**:<br>(No images) / (No times) |

| # | Location City Country | Observer(s) / Equipment | Lat. Lon. Alt. (m) Method | Observations Ingress (C1,C2), Egress (C3,C4) Times in UTC |
|---|---|---|---|---|
| 4 | C.C. Bima Bogotá Colombia | S. Vanegas / DSLR Camera Sony A200, Alt-azimut tripode | +4.809111 -74.039416 2550 GPS | **Ingress**: (No images) / (No times): **Egress**: *Time lapse* / Times C3, C4 (Ph. & Ex.) |
| 5 | – Cundinamarca Colombia | L. Ariza / Telescope LX200-ACF 10" / LPI meade Lunar Planetary Imager | +5.182839 73.626313 2698 GPS | **Ingress**: (No images) / (No times) **Egress**: *Video* / All Times (Ph.) |
| 6 | Observatorio ITM Medellín-Antioquia Colombia | L. F. Ocampo / Canon 1D Mark IV, CMOS APS-H, 16Mpx Schmidt-Cassegrain 10" Meade Telescope | +6.245361 -75.550888 1625 GPS | **Ingress**: (No images) / (No times) **Egress**: *Time lapse* / All Times (Ex.) |
| 7 | – Bello-Antioquia Colombia | J. Gómez H. Cortés / Visual Observations 10x50 Binoculars | +6.307955 -75.567490 1588 GIS | **Ingress**: (No images) / Time C2 (Vi.) **Egress**: (No images) / (No times) |
| 8 | – San Vicente de Ferrer-Antioquia Colombia | J. Ospina EADE ETX 90 EC Maksutov-Cassegrain / Canon Eos 5D | +6.306694 -75.326433 2262 GIS | **Ingress**: Photographs / All times (Ex.) **Egress**: Photographs / All times (Ex.) |
| 9 | – San Diego-Carabobo Venezuela | J. Salas / 10x70 Binocular Lunar Filter | +10.261689 -67.960280 498 GPS | **Ingress**: (No images) / (No times) **Egress**: (No images) / Time C2 (Vi.) |

## 6. Results

After collecting all the observations provided from the 9 different locations we performed two kind of analysis and post-processing tasks.

The first analysis was intended to precisely measure contact times and their uncertainties using photographies and videos taken before, during and after the occultation. Due to local weather conditions in some locations the phenomenon could only be registered for a couple of seconds after the occultation had ended. In other locations, although the occultation was registered during the whole duration of the phenomenon, no pictures or videos were taken at the exact times of the contacts.

For getting precise values of contact times and their uncertainties we devised an interpolation and extrapolation procedure. For that purpose we took two or more pictures close to a given set of contacts (ingress, contacts 1 and 2 or egress, contacts 3 and 4). After aligning the moon accidents in the corresponding set of pictures (see Figure 8), the cartesian position of the Mars disk centroid with respect to a reference point close to the lunar limb, was determined and tagged with the timestamp of the corresponding picture. The cartesian coordinates of the point where lunar limb crosses the occultation coord was also estimated. Observed contact times were calculated using Equation (1) and linear regressions of the centroid cartesian coordinates x, y as a function of time.

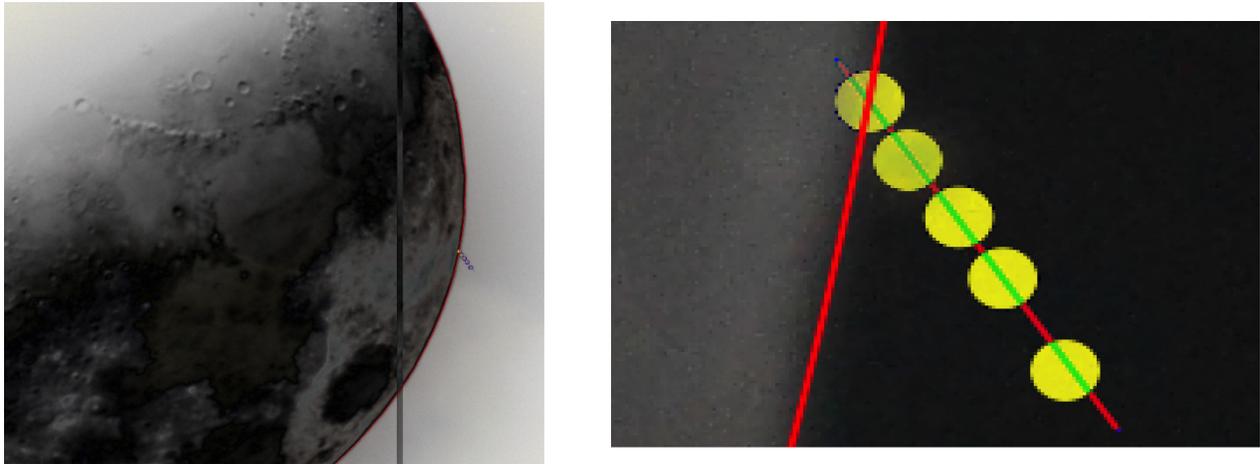

*Figure 8. Extrapolation analysis for the egress pictures taken in Location 8. Left panel: Five pictures showing the Moon and Mars disks have been aligned in such a way that reference lunar accidents coincide in all pictures. Right panel: zoom in of the pictures showing the measured positions of the Mars Disk (yellow disks). The continuous red arc represent the estimated position of the lunar limb. The red and green segment is the occultation cord calculated from the best linear fit of the Mars disk centroids.*

Since the position of the Mars disk centroid is affected by the limited resolution of the images and the uncertainties in the synchronization of each instrument, we used a Monte Carlo procedure to estimate uncertainties in contact times as obtained from the previously described method. Contact times calculated at each location and their respective uncertainties are presented in Table 2. For the sake of completeness we also include there the times of occultation contacts obtained visually at locations 3, 7 and 9.

*Table 2. Measured and estimated contact times at each location. All times are given in UTC-5. Contact times measured visually have been tagged with a "V" prefix. For visual measurements we assumed an equal uncertainty of 10 seconds according to the experience reported by those observers.*

| Location | C1 | C1.5 | C2 | C3 | C3.5 | C4 |
|---|---|---|---|---|---|---|
| 1 | 21:38:39 $^{+11.7}_{-12.7}$ | 21:38:56 $^{+9.2}_{-11.2}$ | 21:39:14 $^{+10.7}_{-11.9}$ | 22:28:18 $^{+8.8}_{-18.0}$ | 22:28:31 $^{+7.8}_{-19.7}$ | 22:28:44 $^{+7.7}_{-18.1}$ |
| 2 | – | – | – | 22:22:6.7 $^{+50.8}_{-72.5}$ | 22:22:20.1 $^{+48}_{-86}$ | 22:22:33.4 $^{+46}_{-94}$ |
| 3 | – | (V)21:05:10 $^{+10}_{-10}$ | – | – | – | – |
| 4 | – | – | – | 22:18:56 $^{+30}_{-30}$ | – | 22:19:49.0 $^{+30}_{-30}$ |
| 5 | – | – | – | 22:17:18.8 $^{+0.8}_{-0.8}$ | – | 22:18:5.7 $^{+0.7}_{-0.7}$ |
| 6 | – | – | – | 22:11:2.6 $^{+7.2}_{-8.9}$ | 22:11:14.9 $^{+6.3}_{-7.8}$ | 22:11:27.2 $^{+6.8}_{-8.4}$ |
| 7 | – | – | (V)21:10:40 $^{+10}_{-10}$ | – | – | – |
| 8 | 21:10:10.2 $^{+25.6}_{-20.6}$ | 21:10:35.6 $^{+22.9}_{-17.6}$ | 21:11:0.9 $^{+26.3}_{-20.9}$ | 22:10:56 $^{+17.3}_{-21.8}$ | 22:11:17.8 $^{+15.2}_{-18.6}$ | 22:11:39.6 $^{+17.5}_{-21}$ |
| 9 | – | – | (V)21:42:39 $^{+10}_{-10}$ | – | – | – |

Values reported in Table 2 show that most of the observations have uncertainties below the light-time effect (~24 seconds). Others were affected by uncertainties too high for this purpose. This was the case of Location 2 contact times that were affected by uncertainties over 1 minute.

Observations from Location 5 seem to provide the most promising measurement to reliably calculate the speed of light. Regretfully we have verified that time synchronization at that location was affected by the largest delay among all sites. Moreover a proper estimation of the exact time-lag between the measured and actual times was impossible after the images were reported to us.

Once we estimated the contact times at each location we proceed at calculating the speed of light that best fit the observed times. For that purpose we first calculate the theoretical contact times $\tau_{ci,j}$ including the light-time effect following the procedure described in Figure 4 and section 4.1. The best fit value of the speed of light is obtained by minimizing the chi-square statistics:

$$\chi^2(c) = \frac{\sum_{i,j} [t_{ci,j} - \tau_{ci,j}(c)]^2}{\Delta t_{ci,j}^2} \quad (1)$$

Here the index $i$ runs over the set of contacts (i = 1,1.5,2,3,3.5,4) and the index $j$, over the locations (j=1-9).

In Table 3 we summarize the results obtained by fitting individually the measurements reported by each team.

*Table 3. Best-fit values of the speed of light as obtained from the set of observations at each location. The last column is the best-fit value obtained by fitting the observations after adding a constant time-lag (value in parenthesis).*

| Location | Ingress Contact Times | Egress Contact Times | All Contact Times | Fitting with a Constant time-lag |
|---|---|---|---|---|
| 1 | 447,477 | 456,010 | 445,593 | 306,611 (-7.0s) |
| 2 | – | 283,832 | – | 302,232 (+1.5s) |
| 3 | X | – | – | – |
| 4 | – | X | – | 293,427 (-98s) |
| 5 | – | X | – | – |
| 6 | – | 306,595 | – | 306,595 (0.0s) |
| 7 | – | 317,188 | – | – |
| 8 | 150,986 | 289,668 | 217,017 | 305,348 (+10s) |
| 9 | X | – | – | 301,240 (+9.5s) |

Individually, most of the measurements reported by the observers are not suitable for estimating the speed of light. The exceptions to this rule are the measurements in Locations 2, 6 and 8 that provides reasonable best fit values for this quantity. Some of the locations seems to be affected by a constant time-lag probably arising from improper synchronization of the clocks inside the instruments.

Thus, for instance, if we correct all the contact times for Location 1 with a constant 7 seconds time-lag, the observations are well fitted with a speed of light close to the actual value of ~300,000 km/s (last column in Table 3). The only exception to this rule are the measurements performed at Location 4 that could not be fitted by including a time-lag correction. This fact reflects that other conditions different than systematic error or bad synchronization, affected the measurements performed at those locations. On the other hand, since only one contact time was measured at locations 3 and 5 a time-lag fitting is useless.

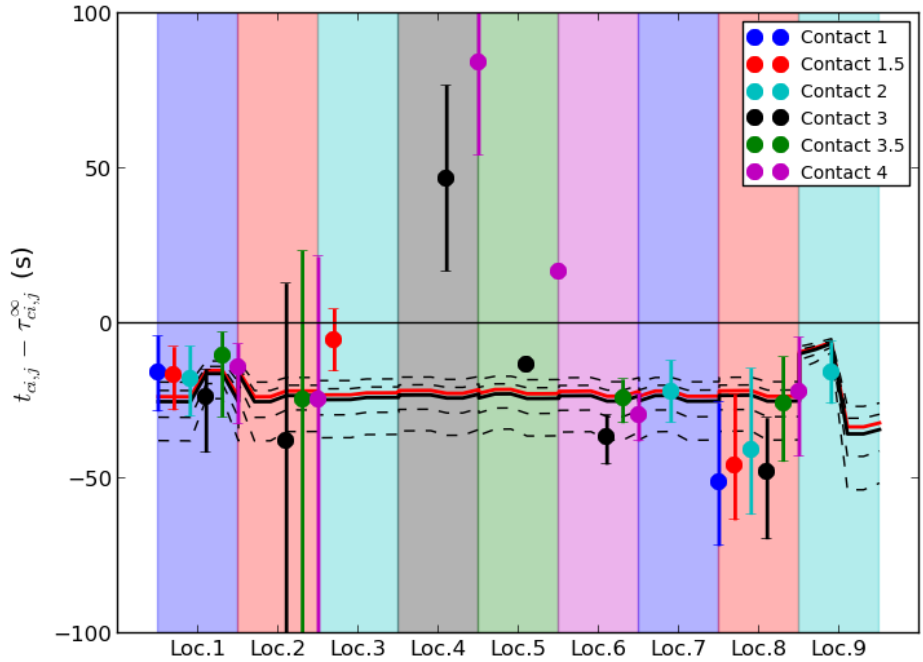

*Figure 9. Difference between the contact times measured at each location and the value expected if the speed of light was infinite ( $\tau_{ci,j}^{\infty}$ ). Dashed lines correspond to the expected contact times difference if the speed of light was (from bottom to top) 200,000, 250,000, 350,000 and 400,000 km/s. The solid black line correspond to the actual speed of light, 300,000 km/s. The solid red line is the best fit value obtained after excluding the most problematic measurements (locations 3, 4 and 5).*

The most interesting result arose when we fit the contact times measured at all Locations. Excluding the problematic measurements performed at Locations 3, 4 and 5, a global fit of the speed of light returned a value of **320,180 km/s** which is off from the true value by just **7%**. This result is pretty robust and varies in a small range when different subsets of locations were included in the fitting procedure.

The second kind of post-processing we performed over the observations was that intended to measure the Lunar Parallax. For that purpose we select those observations having the best quality pictures and the most reliable time measurements. Locations 1, 2, 6 and 8 offered the best opportunities for this analysis.

We first selected a reference time at which the position of the disk of Mars in the reference frame of the Moon could be reliably estimated in most of the locations. We find that a specific time after the egress of Mars at Location 8 was the most suitable reference time. Once the reference time was selected we used pictures at Locations 6, 2 and 1 to extrapolate the position of the disk of Mars behind the Moon. The procedure implied the careful construction of a common reference frame for pictures taken at different locations. For that purpose we select as the origin of the reference frame the crater *Sommering*, easily identifiable in all pictures and located very close to the center of the lunar disk. We also took a common reference direction that was defined using the line joining the center of the Moon and crater *Proculus*, a small and bright crater easily identified in the pictures taken at each location. The scale of the reference axis was set in such a way that the Moon disk has 1,000 units in all pictures, despite their different resolution.

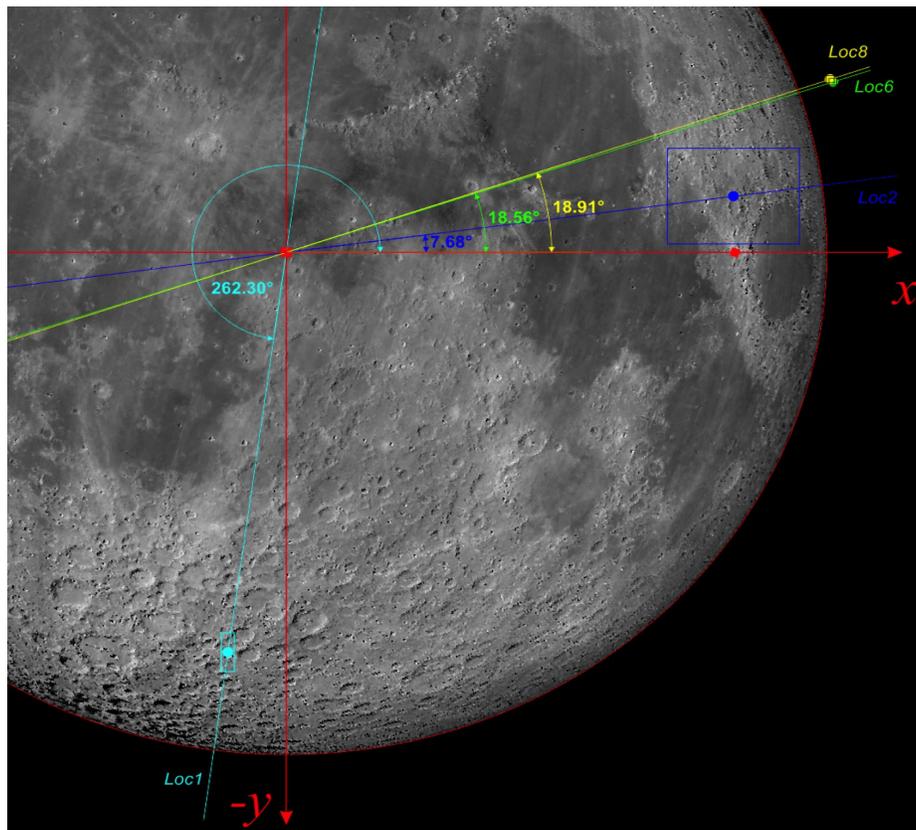

*Figure 10. Lunar reference frame (red lines) and the calculated position of Mars (circles) at a given reference time and it was measured or extrapolated in 4 different locations: Location 1 (cyan), Location 2 (blue), Location 6 (green) and Location 8 (yellow). The parallax of the moon is equal to the angular distance among two of the estimated Mars positions.*

Using a procedure similar to that applied to estimate contact times, we extrapolate the position of the Mars disk at the reference time.

Figure 10 show the position of Mars in the reference frame of the Moon as estimated in 4 different locations. As expected the angular distance between the estimated Mars position for two given locations increases when the geographical distance among the sites is increased (Location 1 is in Chile, Location 2 is in Pasto – Colombia and Locations 6 and 8 are in or close to Medellín – Colombia).

Table 4 summarizes the results of the parallax measurements and first order estimations of the moon distance as obtained from Equations (4) and (5). As expected the best results are obtained when we select the farthest away location as the reference site.

Averaging the three values of the first order lunar distance presented in Table 4 we can conclude that during the Mars occultation of July 15$^{th}$ 2014, the Moon was at a distance 388,000 ± 2,900 km (0.7% uncertainty) from Location 1. This value agrees well with the true distance of 386,909.7 km predicted with detailed theoretical models.

| Reference Location | X | Y | Parallax with respect to location ... | | | |
|---|---|---|---|---|---|---|
| | | | 1 | 2 | 6 | 8 |
| 1 | -108±24 | -796±77 | – | $\pi$ = 0.335258±0.058º<br>B = 2,292.08 km<br>$D_1$ = 391,719.6 km | $\pi$ = 0.410213±0.024º<br>B = 2,768.00 km<br>$D_1$ = 386,617.0 km | $\pi$ = 0.410483±0.024º<br>B = 2,769.76 km<br>$D_1$ = 386,607.0 km |
| 2 | 826±122 | 112±95 | – | – | $\pi$ = 0.07540±0.044º<br>B = 589.13 km<br>$D_1$ = 447,643.2 km | $\pi$ = 0.07586±0.044º<br>B = 604.1 km<br>$D_1$ = 456,281.2 km |
| 6 | 1011±15 | 339±13 | – | – | – | $\pi$ = 0.001730±0.010º<br>B = 25.8 km<br>$D_1$ = 853,409.4 km |
| 8 | 1007±15 | 344±14 | – | – | – | – |

A non-linear fit of the 3 parallaxes to a model where the lunar right ascension, declination and distance are free parameters (see Section 4.3), returned a geocentric distance of 388,484 km which agrees reasonable well with the theoretical expected value of 394,180.5 km within an uncertainty of just 1.4%.

**7. Summary and Conclusions**

In this paper we presented the results of the third Aristarchus Citizen Astronomy Campaign. The Aristarchus Campaigns are intended to involve a large community of observers (amateurs, students and in general citizen scientist) in observational campaigns aimed to reproduce classical astronomy observations and/or to measure well known astronomical and physical constants.

The final goal of the campaign is that amateurs and astronomy enthusiasts in developing countries recognize that by using simple instruments, including readily available electronic gadgets, they can contribute to measure collaboratively the local Universe. Training the communities of amateur astronomers in the developing world for participating in advanced observational campaigns could be very beneficial for the advancement of Astronomy in general.

We have used observations submitted from 9 different locations in the North part of South America of the Mars occultation by the Moon in July 5$^{th}$ 2014, to estimate the speed of light and the lunar distance.

Using a method that exploits the fact that light-time affects the precise time of occultation contacts and combining most of the observations gathered by the collaborators we obtained a best-fit value of 320,180 km/s which is only 7% different than the exact value of this constant. We also used the fact that some of the observations were made with base-lines as large as 2,500 km to calculate the lunar distance. Our result had an uncertainty of just ~1%, almost 3 times more precise than other simple methods.

The formula and methods devised here can be used in different contexts. Astronomy and Physics students can exploit the examples provided by this work. Since occultations of the brightest planets by the moon are relatively common, this exercise could be repeated regularly until reaching more precise results.

For those attempting to repeat these measurements we provide source codes in C and python able to calculate the physical and astronomical quantities required for the analysis. The codes can be obtained from the `GitHub` repository http://github.com/facom/iContact.

When repeating these measurements and analysis, be sure to take care of the following details:

- Contact times can be measured with exquisite precision if instead of pictures you take videos and record an audible time signal. Free software can be obtained in Internet where a connection with a time server converts the digital time signal in analogue sounds that can be recorded by the camera.

- If using cameras and pictures, the best way to properly record time is taking pictures of the computer screen including a time signal on Internet using the same camera and at almost the same time when pictures are taken.

- Using the highest resolution in the camera as well as raw format will produce the lowest uncertainties in any measurement procedure over the images.

- Distances larger than 1,000 km are desirable to obtain really precise estimations of the lunar distance. For that purpose contacting people as far as possible from your location is almost mandatory if you want to exploit this possibility.

- Use GPS and cellphones instead of GIS software as Google Earth when determining your geographical position in this kind of campaigns. Moreover, be sure that the GPS and cellphone is operating outdoors and that the maximum number of satellites are providing information to the equipment about your position.

**Acknowledgements**

We want to thank to those friends and colleagues who support us through messages in social networks encouraging people to participate in the Aristarchus Campaign. The Campaign is an initiative of the Sociedad Antioqueña de Astronomía (SAA) and University of Antioquia. We thank both organizations and institutions for promoting and supporting this kind of public initiatives. Special thanks to Joaquin G. Ossa who participated as photographer at Location 6. Jorge I. Zuluaga is funded by Vicerrectoría de Docencia/UdeA, CODI and the Fulbright Commission Colombia. JIZ also thanks to Harvard-

**Appendix A. Precise formulas for determining the lunar distance from the lunar parallax**

The formulas developed here refer to the geometrical construction in Figure 4. Let's assume that we know the value of the lunar parallax $\pi$, and the celestial coordinates of the Moon center $(\alpha, \delta)$ as measured from Location 1. The unitary vector towards the Moon is:

$$\hat{u}_M : (\cos\delta \cos\alpha, \cos\delta \sin\alpha, \sin\delta) \quad (A1)$$

Using the geographic coordinates of the observing sites $(\lambda, \phi, h)$ and the parameters of the reference Earth ellipsoid (equatorial radius $a$ and flattening $f$), the geocentric position vector $\vec{r}_{EO1}$, $\vec{r}_{EO2}$ of the observers can be precisely determined using the following prescription:

$$e^2 = f(2-f)$$

$$R = \frac{a}{\sqrt{1-e^2 \sin^2\varphi}}$$

$$\begin{aligned} x: & \quad (R+h)\cos\varphi \sin\lambda \\ y: & \quad (R+h)\cos\varphi \cos\lambda \\ z: & \quad (R[1-e^2]+h)\sin\varphi \end{aligned} \quad (A2)$$

$$\vec{r} : (x, y, z)$$

Finally the vector pointing from Observer 1 to Observer 2, $\Delta\vec{r}_{O1O2} = \vec{r}_{O2} - \vec{r}_{O1}$, complete the set of quantities required to determine precisely the lunar distance. For that purpose we require the angles $\theta$ and $\beta$ at each observing site (see Figure 4) that can be obtained from the dot products among the previous defined vectors:

$$\cos\theta_1 = \frac{\Delta\vec{r}_{O1O2} \cdot \hat{u}_M}{B}$$

$$\cos\beta_1 = \frac{-\Delta\vec{r}_{O1O2} \cdot \vec{r}_{O1}}{R_1 B} \quad (A3)$$

$$\theta_2 = 180 - \theta_1 - \pi$$

$$\beta_2 = 180 - \beta_1 - \Delta$$

Where $\Delta$ is the great circle distance among the observing sites. The lunar distances can be finally obtained from the application of the sine and cosine law to several of the triangles in the construction:

$$D_1 = B \frac{\sin \theta_2}{\sin \pi}$$

$$D_2 = B \frac{\sin \theta_1}{\sin \pi}$$

(A3)